# The "dark dips" phenomenon in the LSST Camera on-sky images

Claire Juramy*[a], Pierre Antilogus[a], Pierre Astier[a], John Banovetz[b,c], Sean Patrick MacBride[d], Andrew P. Rasmussen[e], Yousuke Utsumi[f]

[a]Laboratoire de Physique Nucléaire et des Hautes Énergies, Sorbonne Université, Université Paris Cité, CNRS/IN2P3, 4 place Jussieu, 75005 Paris, France; [b]Lawrence Berkeley National Laboratory, 1 Cyclotron Road, Berkeley, CA 94720, USA; [c]Physics Department, Brookhaven National Laboratory, Upton, NY 11973, USA; [d]Physik-Institut, University of Zurich, Winterthurerstrasse 190, 8057 Zurich, Switzerland; [e]SLAC National Accelerator Laboratory, Menlo Park, CA, USA; [f]National Astronomical Observatory of Japan, 2-21-1 Osawa, Mitaka, Tokyo 181-8588, Japan

**ABSTRACT**

When the commissioning camera (ComCam), and then the LSST Camera, started taking on-sky images at the Vera Rubin Observatory, some of the ITL STA3800 CCDs exhibited a previously undocumented effect. When a sufficiently bright star is superimposed over the sky background, the sensor columns that contain the star appear slightly darker than the background, both above and below the position of the star. The visual appearance of these 'dark dips' in the background is enhanced by a slight excess of flux within the neighboring columns, suggesting that they are caused by a lateral field distortion that shifts charges away from the central columns.

This effect is not uniform across sensors, reaching an amplitude of up to seven percent of the background in the worst cases, but being undetectable in other sensors under the same conditions. For the affected sensors, the threshold required for the dips to become detectable also varies, from a few hundred to a few thousands saturated pixels within the star footprint.

We developed means to quantify the dips, and studied the effect statistically over hundreds of frames, in order to categorize each sensor. In particular, we found no significant dependence on the filter color. We then developed a strategy to dynamically mask the affected columns during the Instrument Signature Removal process, to avoid any potential effect on photometric and astrometric performance.

## 1. INTRODUCTION

The LSST Camera of the Vera C. Rubin Observatory is a composite focal plane, with a science array of 117 Teledyne-e2v CCD250 and 72 ITL STA3800C devices, and 12 more ITL devices for guiding and focus. Aside from their basic physical properties (10 x 10 µm pixels, 100 µm thickness, divided into 16 channels of approximately 1 million pixels each), the designs and the manufacturing processes of those devices differ significantly. The operation of these devices required to be adapted accordingly [1], and various challenges encountered along the way to eliminate problematic effects were also manufacturer-dependent [1, 2]. The "dark dips" effect that we will describe in this paper is one such effect, since it affects only some ITL sensors. It remained undetected during several testing and characterization runs that were performed before the camera went on sky. This is partly due to the previously undocumented nature of this phenomenon, but also to the specific conditions that are required to reveal it, which are a combination of a strongly saturated spot, and a sufficiently high background level to reveal anomalies at the percent level.

Once the commissioning camera (ComCam), composed of one "raft" of nine ITL sensors, started taking images on-sky, the effect was quickly noticed (see Figure 1). We then attempted to reproduce it on the main camera, which was undergoing its final round of tests at the Observatory, but were unsuccessful [2], due to the limited test hardware available. The effect was then studied and characterized extensively on ComCam, across all filters, and on a spare ITL raft at SLAC. Once the main camera went on sky, the same processing was applied to all ITL sensors, to determine to which extent they were affected.

*juramy@lpnhe.in2p3.fr

In this paper, we will first describe how we detect and measure the “dark dips” in on-sky images. Then we will show the results of a statistical study conducted over all ITL science sensors of the LSST Camera, quantifying the frequency of the effect, but also the variations in how it presents across sensors. We will present some dedicated studies that were carried out to try to better understand the origin of the phenomenon, and some attempts that were made to mitigate it, concluding with the current implementation in the LSST pipeline.

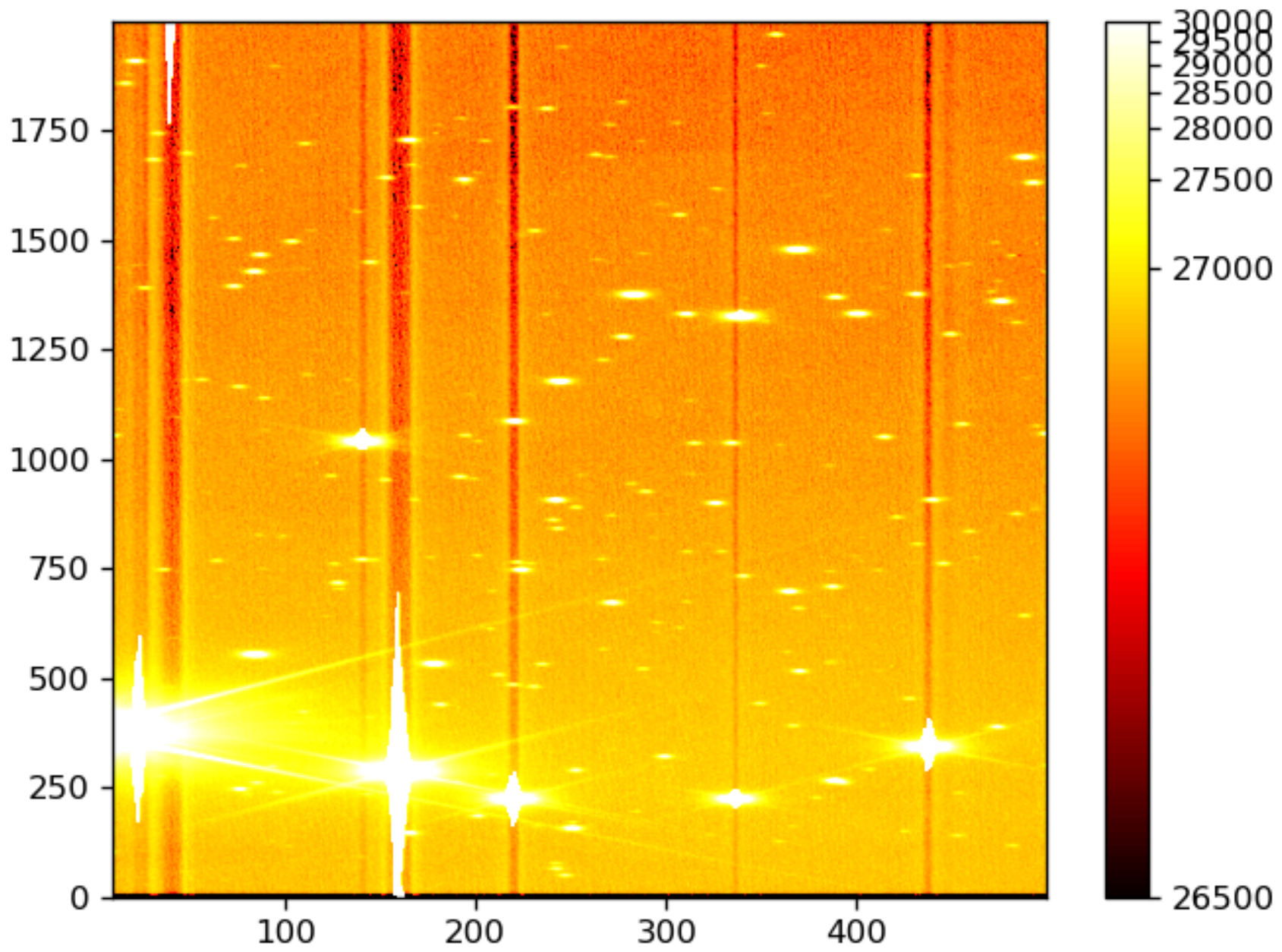

Figure 1. The “dark dips” issue in a segment of ComCam sensor S22 (509 columns, 2000 rows). This is a raw image in digital units, without any processing. All sets of darker columns are associated with a bright star, with some coming from the segment directly above this one (not shown). The edges of the wider dips are slightly brighter than the surrounding background.

## 2. Processing for the dark dips

This section follows the four steps of processing for the “dark dips” over a large number of frames. This process was done for characterization purposes, and is distinct from the implementation of dip mitigation in the Instrument Signature Removal (ISR) process, which we will describe in a later section.

### 2.1 Find clusters of saturated pixels

The Instrument Signature Removal is the first processing step of the LSST pipeline [3]. It converts the raw numerical data coming out of the data acquisition system into full-sensor frames, expressed in electrons. Along the way, the ISR generates a set of pixel masks that flags various issues. One such flag is the ‘SATURATED’ flag, which tags all pixels above a threshold (as defined in [2]). In effect, this flags not only the bright object, but also the pixels contaminated by the bleeding charges when the object is sufficiently bright. The first step is therefore to scan the saturated mask for each frame, and group all adjacent pixels into ‘clusters’. The coordinates of each cluster are recorded (approximate center, and the extent over columns and rows), and the total count of saturated pixels. A side product of this search is that we chart the growth of the saturated clusters as their counts increase, both in width across columns and length across rows (see Figure 2).

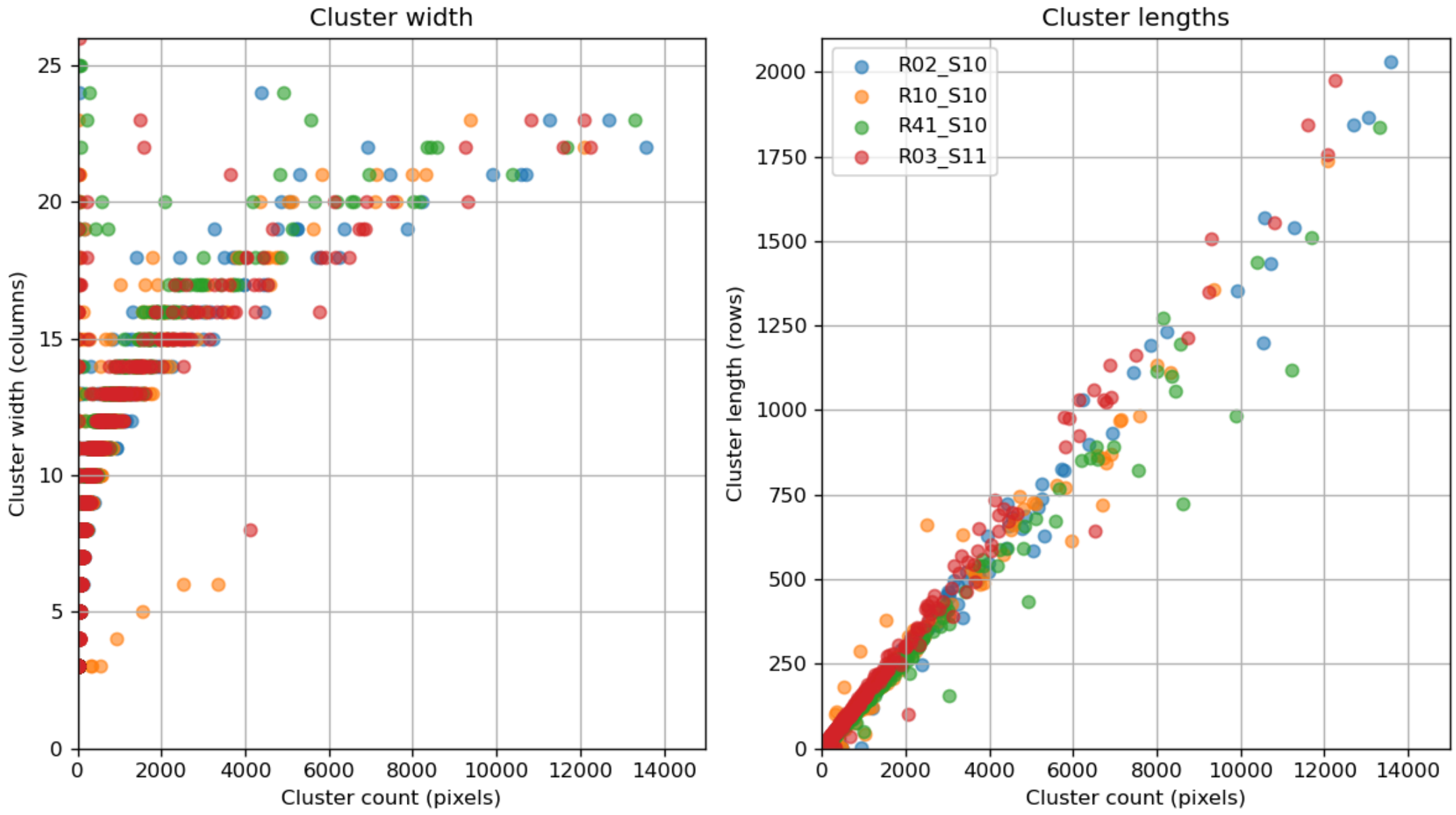

Figure 2. Increase in width over sensor columns and bleeding over rows as the number of saturated pixels increases in a cluster, for a set of 100 ITL frames in the i-filter. The four sensors were selected to sample the range of behavior of the dip phenomenon: R02_S10 is strongly affected, R10_S10 is only affected by larger clusters producing wide but shallow dips, R41_S10 is affected at similar levels but with more narrow dips, and R03_S11 is not affected at all.

### 2.2 Compute the median and fit a dip profile

For each cluster of saturated pixels, the per-column median is computed separately over the areas above and below the saturated cluster, within the cluster and in the adjacent 'aisles' (30 columns on each side). The equation for the dip fit $f$ as a function of column coordinate $x$ is:

$$f(x) = a(x-x_0) + b + (r(x-x_0)^2 - c) \,.\, exp(-(x-x_0)^2/w^2)$$

With:

- $x_0$ the coordinate of the center of the dip
- $a$ the slope of the background (required as the sky background is not necessarily uniform)
- $b$ the constant component of the background
- $c$ the constant coefficient of the gaussian dip (referred to as 'dipping')
- $r$ the 'rim' factor that models the excess on the sides of the dip
- $w$ the gaussian width of the dip

All parameters except the slope are bound to be positive or null. This function was built up empirically, to fit the data with the minimum number of parameters, with the focus on being able to match the dip shape even from noisy data with only a few points within the cluster footprint. While the central gaussian accounts fairly closely for the depth and width of the darker part of the dip (see Figure 3), the areas of excess flux around the dip, when they are present, are only roughly accounted for by the 'rim' parameter. Figure 3 showcases the fit results in a few typical configurations.

In addition to the fit parameters, the top-to-bottom depth of the dips is estimated by computing the minimum and the maximum of the fitted function at the discrete pixel positions within the cluster footprint. A local value of the noise is also computed as the standard deviation of the residuals to the fit, limited to within the 'aisles' outside of the cluster footprint. This estimates local random variations for the next step, to check the significance of the fitted dip.

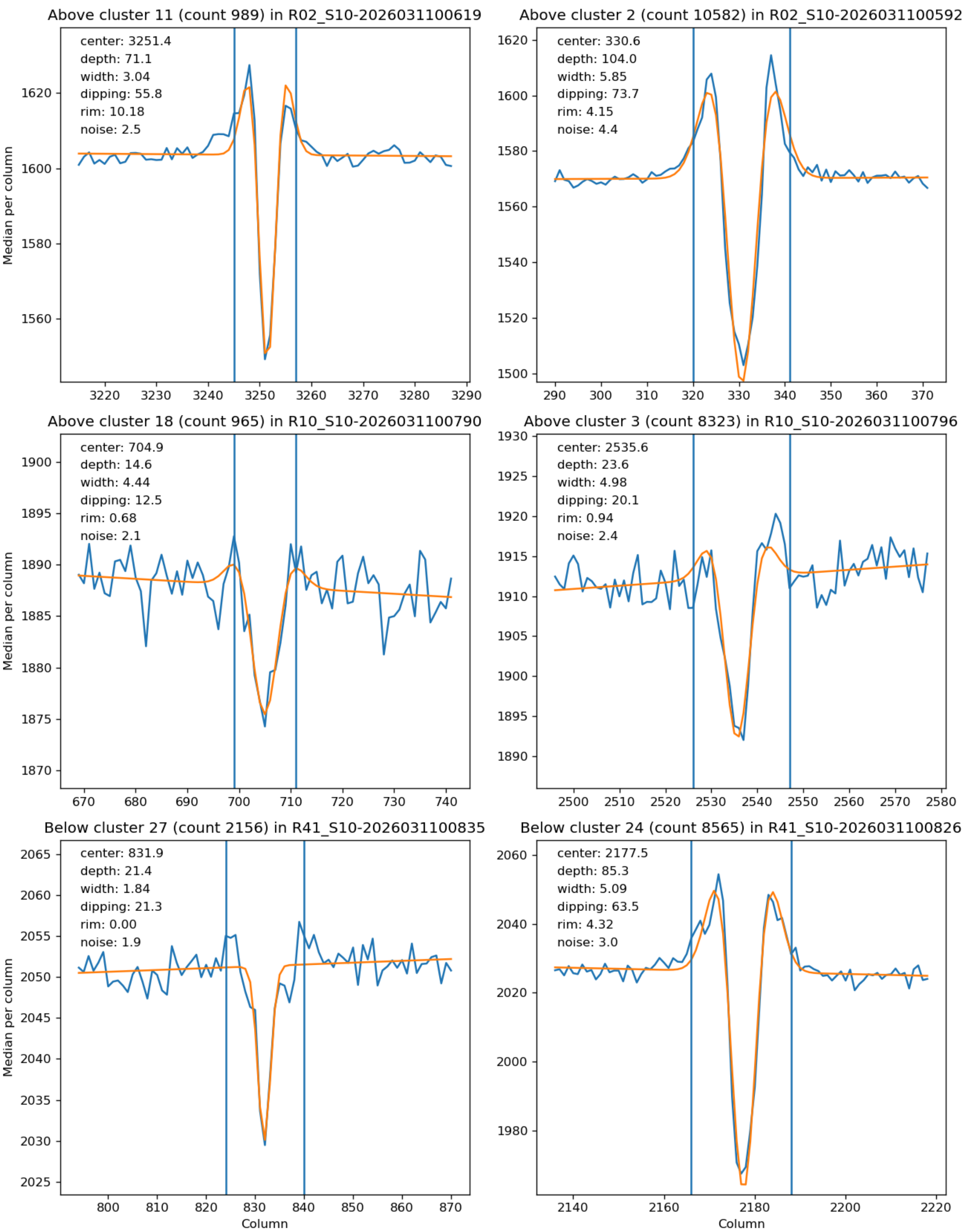

Figure 3. Some typical dips, with the median over columns (in blue), the fitted function at the discrete pixel positions (in orange), and the parameters resulting from the fit for each dip. The blue vertical lines mark the boundaries of the saturated cluster that is associated to the dip. The fit results illustrate the variability of the dip behaviors: very strong dips even at lower cluster counts for R02_S10 (first row), wide but shallow dips for R10_S10 (second row), narrow dips at low cluster counts that get wider at high counts for R41_S10 (third row).

### 2.3 Determine which clusters have a detectable dip

The fitted parameters for each cluster are filtered, to determine if they match the criteria that we set for a dip to be detected. This is still done separately for the top and bottom fit results, and a detection is obtained if either side meets all the criteria. The criteria that we used for detection are listed in Table 1. Setting those values was a trade-off between triggering detections on random background noise, and missing real dips. The requirement on width is the minimum value to avoid triggering on random spikes, and the requirements on the depth relative to the noise ensure that the dip is indeed significant. Other criteria are sanity cuts that were added from experience, to cut out random features passing off as dips. On balance, our main concern was not to miss that a sensor could exhibit dips, which would have compromised the quality of the survey images. Accordingly, the finalized values were set as low as possible, while still filtering most noise and random features.

Table 1. Criteria for dip detection.

| Parameter | Criterion |
|---|---|
| Dipping (constant of the gaussian) | > 3 * Noise |
| Depth (top-to-bottom) | > 4 * Noise |
| Depth (top-to-bottom) | < Background / 2 |
| Width | > 1.75 |
| Center ± Width * 0.9 | Within footprint of saturated cluster |
| Rim factor | < Depth |

### 2.4 Check for missed and spurious dips

Before examining the results on a case-by-case basis, there are a few obvious sources of fake dips that can be filtered out. The first are bright defects within the sensors, which can imitate a dip when the defect also creates a darker trail in the later rows. Those are easily detected, because they always appear at the same coordinates, with similar counts (at the nominal 30 s exposure time). Another is a coincidence between two (or more) bright clusters that touch the same columns: if the bigger cluster creates a significant dip, it will also be associated with the smaller cluster. Every cluster is checked for such coincidences, and the smaller cluster(s) are tagged, to remove them from later statistical studies. Lastly, the background level drops near the edges of our sensors, so a 15-pixel wide area is excluded at all edges.

To assess the adequacy of our chosen cuts, we performed a number of checks on individual cases. To address the concern about correctly classifying sensors, the first check was aimed at uncovering potential missed dips, focusing on the clusters with the highest counts (excluding those that were saturated across the whole sensor). We found no case of a sensor in which dips were missed entirely. Occasionally, dips may be missed within sensors that are very sensitive to the effect, when a larger dip overlaps the ‘aisle’ area of a smaller dip, increasing the local noise threshold.

The second angle was to check for missed dips at the lower counts, for sensors that exhibited dips at higher counts. We found that, below certain counts, a fraction of the dip candidates in those sensors become indistinguishable from the noise.

The last check was performed on all sensors that showed a statistically very low number of dips over the sample frames. We verified that those few detections were spurious, and that those sensors could in fact be safely classified as not affected by the dip phenomenon. One source of spurious dips occurs in frames with a high density of objects, in which the random placement of objects can rarely generate a dip detection. This affected particularly frames in the z filter, in which the density of bright objects was higher than in the i-filter. There were also a few cases in which bad background subtractions imitated shallow dips, due to large saturated objects hitting across the middle of the sensor, and bleeding into the parallel overscan that is acquired after each frame.

## 3. Statistical variations across sensors

The results and plots in this section are based on a specific sample of 100 frames acquired in the i filter on March 11th, 2026, but the results are compatible with tests conducted over multiple other samples. To illustrate how affected each sensor is, the number of dips that were detected is compared to the total number of clusters of saturated pixels (see Figure 4). This comparison is focused on a selection at the higher ranges of cluster size (above 3,000), while cutting off saturated objects that traverse the whole sensor at higher counts (above 20,000). After eliminating spurious detections, there are 30 sensors out of 72 that are definitely affected by the "dark dips" phenomenon, with 2 more which exhibit dips very rarely, and only in association to very bright objects.

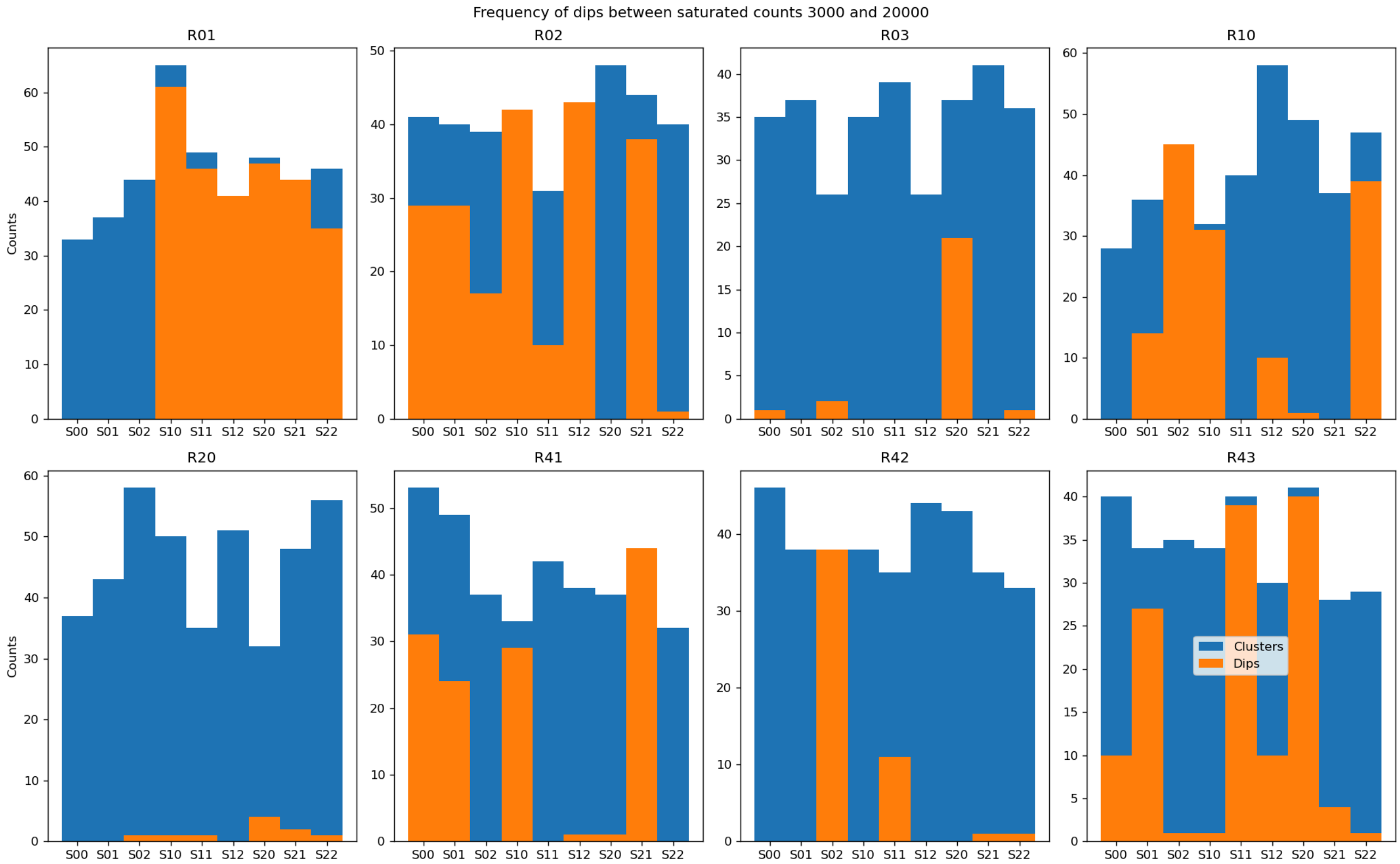

Figure 4. Number of detected dips for all ITL sensors of the LSST Camera. In blue: counts of saturated clusters within the 3,000 to 20,000 size range, for 100 i-filter frames. In orange: counts of detected dips for the same frames. Cases with only 1-2 detected dips can safely be classified as spurious detections, due to noise or background features. Sensors with more than 2 detected dips need to be considered as impacted by the effect, to varying degrees.

To get a more complete picture of the extent to which each sensor is affected, we can look at the frequency of dips, split by bins of cluster size (as in Figure 5). The threshold required for the onset of the dip phenomenon varies strongly across sensors. The variability between sensors applies also to the distribution of dip parameters, as illustrated over a representative sample of sensors in Figure 6. Over all sensors, the negative component of the dip reaches a maximum of 6% of the background level, and the full top-to-bottom depth peaks at 8%, a figure which may be slightly underestimated due to the limitation of the fit of the dip rim. From these distributions, we built up estimators to classify each affected sensor according to the onset of the dips, and in the depth-width plane (Figure 7). The distribution of all those parameters matches qualitative observations, that only 12 sensors are highly sensitive to the effect. These can be defined as the sensors in which dips become detectable at saturation counts below 500, and are also the sensors which exhibit the deepest dips. The rest of the sensors follow a trend towards higher thresholds and shallower dips, including roughly a half of affected sensors that stay in the 1% depth range.

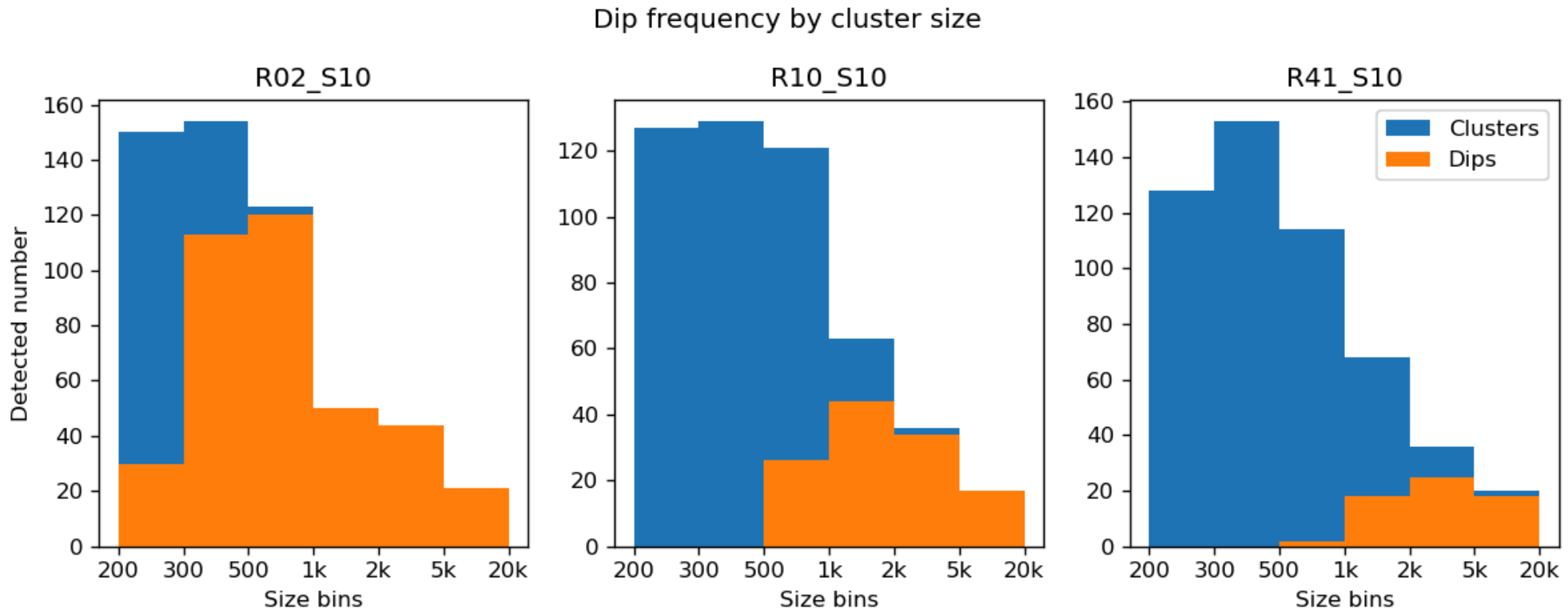

Figure 5. Number of detected dips (in orange), relative to the number of saturated clusters (in blue), for our three representative sensors. The bins are split by total count in the saturated clusters. R02_S10 is among the sensors most affected by the dips, with detectable dips appearing as early as the 200-300 counts bin.

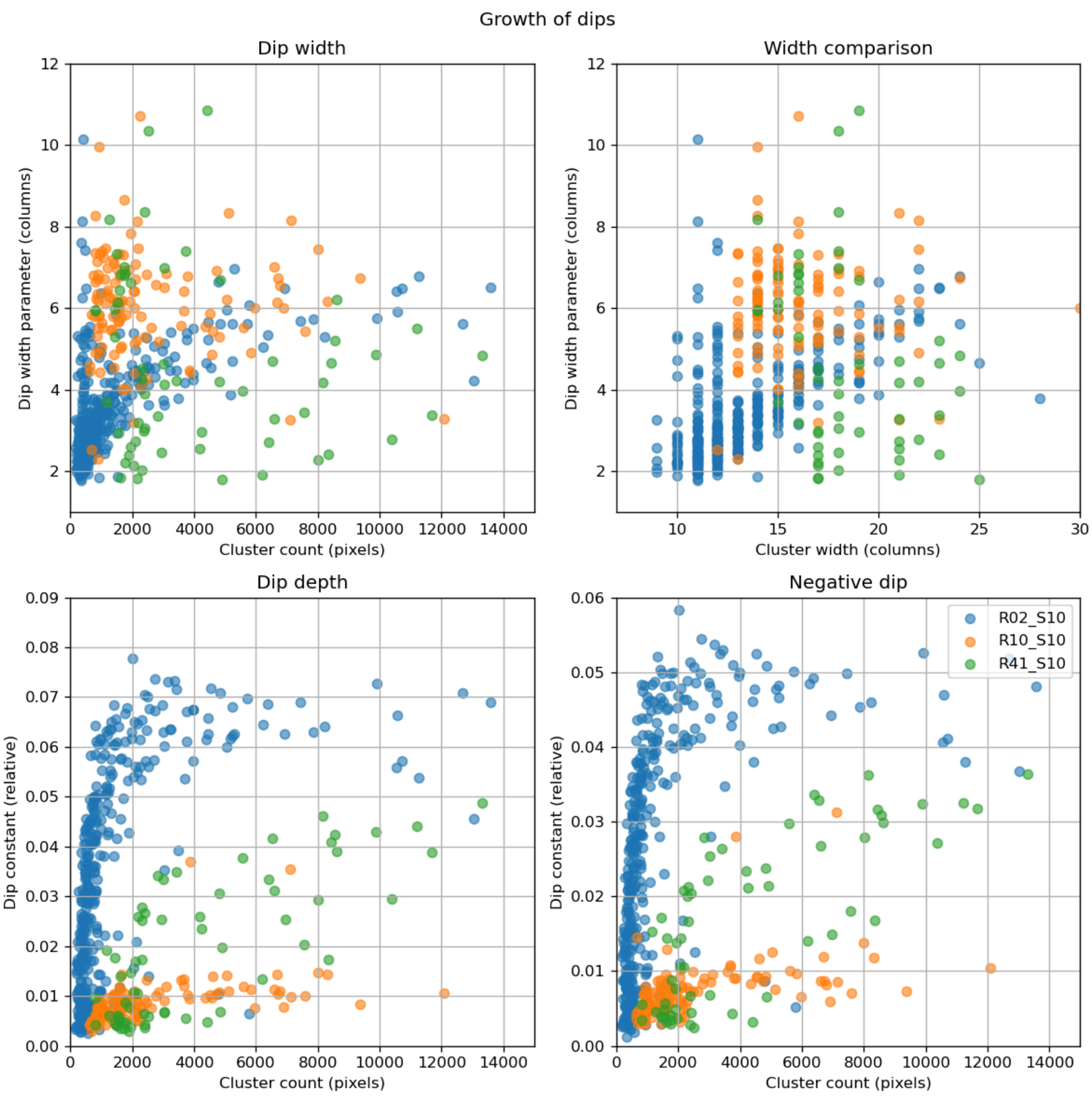

Figure 6. Distribution of dip parameters over 100 frames, depending on cluster size, for our three representative sensors. Top row: dip width, as a function of cluster count (left) and cluster width (right), showing in particular the difference between R10_S10 (wider dips) and R41_S10 (narrower dips). Bottom row: total depth (left) and negative part of the dip (right) as a function of cluster size, showing the rapid growth of the dips at low cluster counts for R02_S10, and the shallowness of dips for R10_S10.

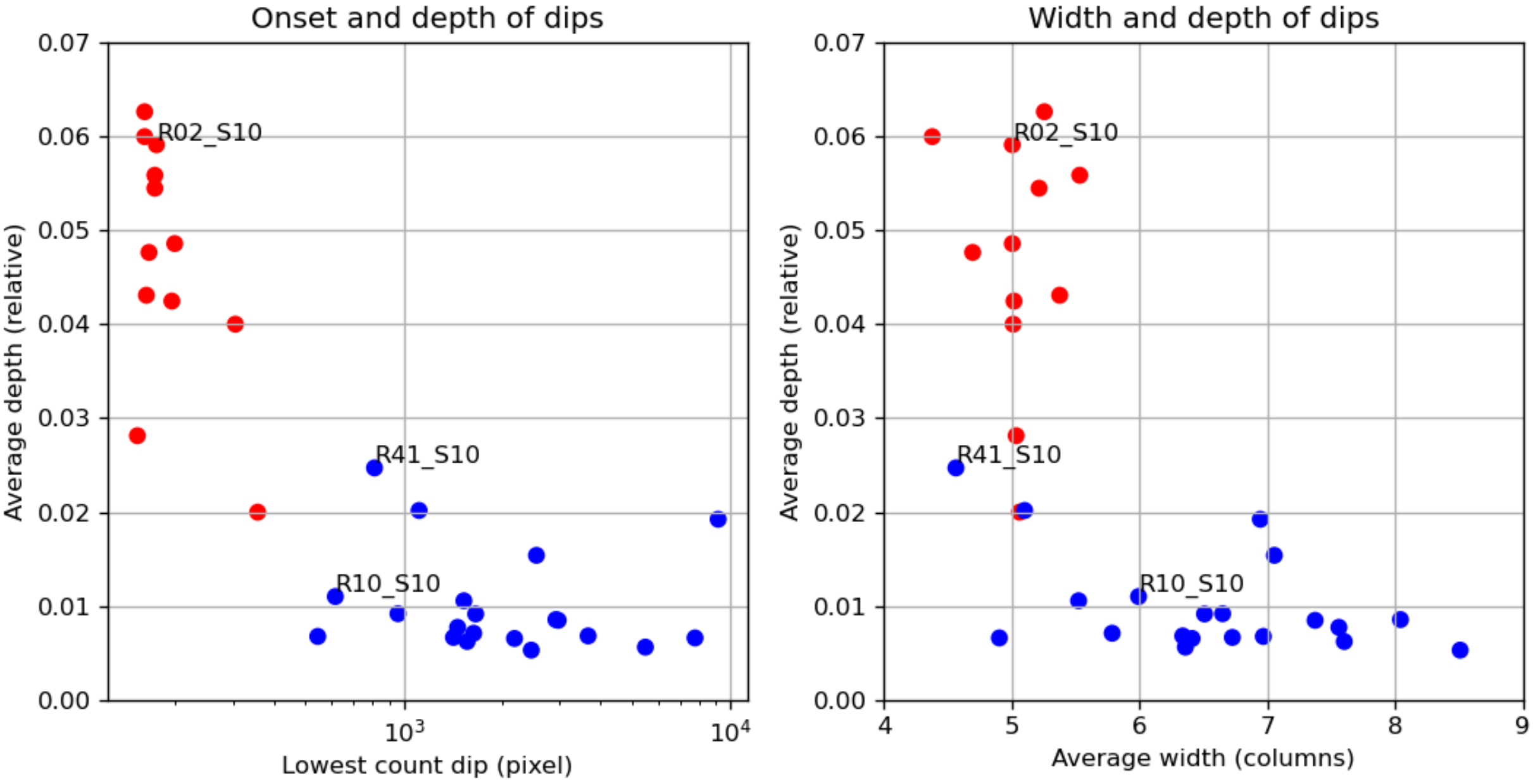

Figure 7. Distribution of all sensors affected by dips, highlighting (in red) the sensors that are strongly sensitive to the dip phenomenon. Left: relation between dip onset (counts in the cluster) and depth. Right: distribution in the width-depth plane. The depth and width parameters are averaged over a range of cluster sizes of 2,000 to 15,000, based on the distribution in Figure 6.

One other question that we wanted to address with large samples of dips was whether there was a spatial dependency of the phenomenon across individual sensors, which could give us hints as to its nature. Mapping the positions and amplitudes of the dips across sensors did not reveal any significant pattern in their distribution. There were however hints of an asymmetry between the dip depths towards the top and the bottom of the sensors, depending on the position of the cluster. This is confirmed by plotting specifically the difference between those two quantities, as a function of the row position of the center of the cluster (Figure 8). The average of this difference varies by 0.5% of the background level over the extent of the sensor, with the dips being slightly deeper towards one side when they are closer to it.

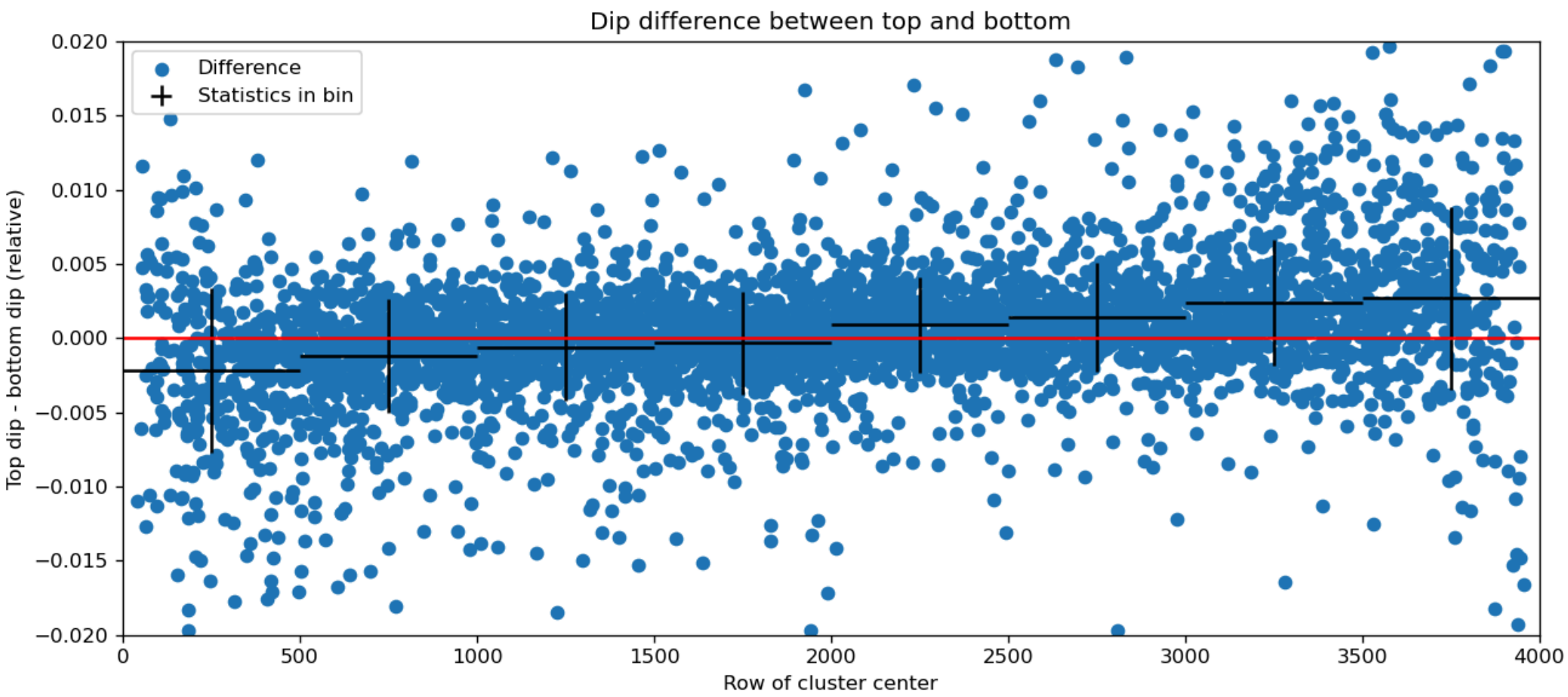

Figure 8. Difference between the negative components of the fit result above and below each cluster of saturated pixels, plotted against the center row coordinate of the cluster, for all detected dips in a sample of 100 frames. In our coordinates, row 0 is defined as the ‘bottom’ side, 4000 as the highest row.

## 4. Qualitative studies

A wide array of studies was conducted during the LSST commissioning, with ComCam, in the lab at SLAC and at Vera Rubin Observatory, and finally with the main Camera on the sky. Beyond assessing the fallout of this issue, we aimed to understand its origin, which might suggest pathways to remedy it. The fact that the dips are present both before and after the source object in the CCD readout, not to mention in the segment on the opposite side of the sensor, make it clear that this is a phenomenon that occurs within the sensor, not during the readout. The association between the saturated object and the dip means that this occurs during the exposure time, and that through some unknown mechanism the saturation in one specific area of the sensor affects the distribution of the electrons in the pixels elsewhere.

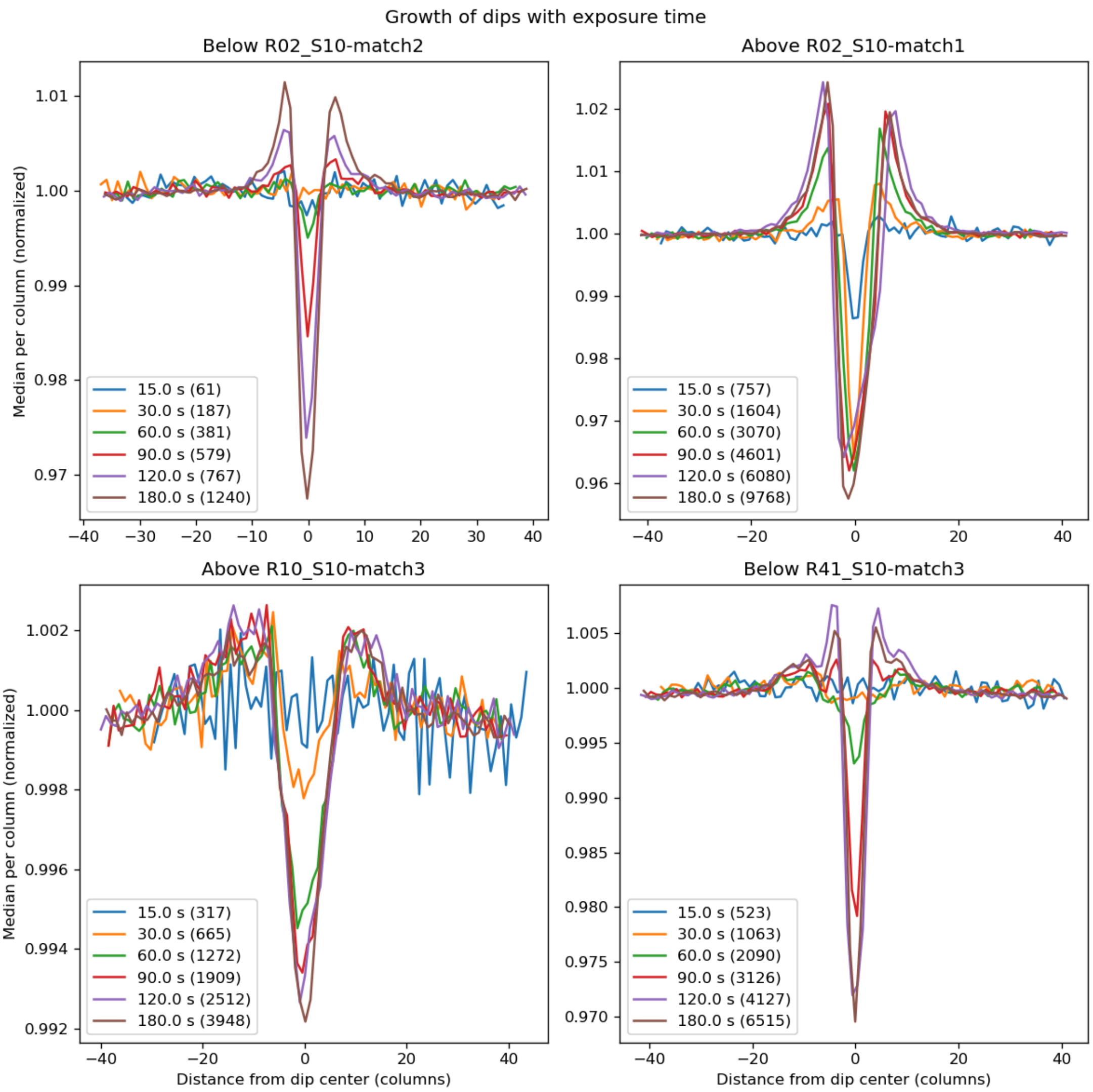

Figure 9. Dips for the same bright object at varying exposure times (in parentheses, the count of saturated pixels for each time), relative to the background level. Top left: increase in dip depth for a low-count cluster in sensor R02_S10. Top right: widening of the dip for a high-count cluster in the same sensor. Bottom left: emergence and increase in depth of a wide dip in R10_S10. Bottom right: emergence and increase in depth of a narrow dip in R41_S10.

### 4.1 Benchtop tests

After a first unsuccessful attempt to reproduce this effect during the last test run with the LSST Camera [2], we were able to assemble a setup that could project separately a bright point source and a flat field at SLAC, and to use it on a sensor that exhibits the dip effect at a 1% level. By separating the two phases of illumination, we were able to confirm that the dip only occurs if the flat field is projected onto the sensor after it has been exposed to the bright point source, and not before. Adding a 40 s delay between the spot and the flat field resulted in a relative decrease of the dip depth by 20%, suggesting the existence of a slow decay path. Tests on this setup also showed that the dips did not appear when the back-substrate polarization was turned off. However, it should be noted that when this voltage was turned off, the charges of the bright spot were spread out over a wider area, and did not exhibit bleeding behavior along sensor columns.

### 4.2 Temperature and color dependency

While the working point of the LSST Camera has been extremely stable over its operation, this was not the case with ComCam, which acquired data with CCD temperatures ranging from -100˚C to -75˚C. There was no detectable difference in the amplitudes and frequencies of the dips over this range of temperatures.
ComCam data also allowed us to verify early on that the dips were present, and had a similar distribution of parameters, in five filters of the LSST survey (g, r, i, z, Y). In the case of the u filter, it was impossible to confirm the presence of the dips due to the low level of the background. The filter-dependent study was later confirmed with data from the LSST Camera, with the same limitation that in bluer filters, shallow dips are harder to detect due to lower background levels.

### 4.3 Dedicated dip measurements

One early source of concern was whether the dips that appear in one frame could carry over into the next, even after the saturated object goes away. A study was carried out on ComCam data, and repeated with LSST Camera data, using the dip detection pipeline over the next frames in the same spots as dips detected in the previous frames. No persistent effect was found.
One short dedicated study was conducted with the LSST Camera on sky: a series of frames was acquired in the i filter, tracking the same area of the sky, and varying the exposure times from 0.1 to 180s. This allowed us to reconstruct the growth of the dips as the saturated objects accumulate charges during an exposure (see Figure 9). In the case of a sensor that is highly sensitive to dips, we see the increase of the depth after the dip first emerges. Then, as the saturated count increases, the dip depth reaches a limit, and the dip widens, while the flux excess moves away from the dip center. For sensors that are less affected, we see the emergence of the dips at higher counts, and a non-linear increase in dip depth, which remains much lower overall.

## 5. Possible mechanism of the “dark dips” and Mitigation

From our studies, we can draw a fairly clear empirical picture of the “dark dips” phenomenon. However, to explain it requires a mechanism that can dynamically alter the distribution of charges in the pixels based on the overflowing of charges elsewhere along the same sensor columns. This could be happening either while the charges are drifting from their conversion point to the pixel potential well, or once they are in the pixel potential wells.

### 5.1 Drift field distortion

The excess of flux around the darker center columns is in favor of the first model, since it suggests that a fraction of the charges is shifted from the center columns to the outer columns. As in the case of the brighter-fatter effect [4], this could be the result of a local electric field that modifies slightly the original drift field lines, changing the effective collection areas of the pixels. However, the brighter-fatter effect is limited in range to a few pixels neighboring the center pixel in which the charges are accumulated.
Another case of local field distortion is the “tearing” that was observed under some conditions in the Teledyne-e2v sensors of the LSST Camera [5]. In particular, it can present as a “divisadero” effect [1], which appears characteristically as a pair of darker columns marking the edges of each of the 16 segments (excluding sensor edges), surrounded by an excess in the flat field flux that decreases slowly when moving away from the edges. The maximum amplitude of this distortion of the flat field is around 3% [5]. The explanation of the “divisadero” is not found in the accumulation of charges in pixels, but in the depletion of hole carriers in the channel stop at the edge between segments. This depletion over the whole length of the channel stop makes the two neighboring sensor columns attract fewer of the incoming electrons during the exposure, by creating a shift of the drift field away from those two columns and into those further away. No similar effect was observed in the ITL sensors of the LSST Camera, which was attributed to the differences in operating voltages [1]. The process used for implanting the channel stops is also very different between the two designs [6].

Due to the similarities in shape and amplitude of the "dark dips" to the "divisadero", the channel stops were considered good candidates as sources of drift field distortions. Under that hypothesis, we attempted to remediate the "dark dips" in the same way that we can remediate the "divisadero": by modifying the state of the serial clock phases during the exposure. In the case of the "divisadero", it appears that we are able to block the holes within the edge channel stops by keeping the serial clocks high. However, testing various combinations of serial clocks (low, high, shifting, fixed) during the exposures had no effect on the presence and amplitude of the dips.
We are also missing a mechanism through which the electrons bleeding along sensor columns could affect the channel stops from the top to the bottom of the sensors, without being impeded by the barrier potentials of the parallel clocks. One other major objection to explaining the dips with a drift field distortion is that these effects should conserve the total charge, which is not the case in our dip profiles. However, while using the median is practical to compute a profile and measure the dip parameters, it might not account fully for the charge distribution in the sides of the dips.
The alternative to an explanation based purely on drift field distortion would be that reaching bleeding condition within one part of a pixel column affects the rest of the column, possibly leading to the loss of charges during the exposure. This phenomenon only manifests once the bleeding reaches a significant extent, at which point we are outside of the conditions underpinning the expected behavior of potential wells and buried channels [7]. Through the usual mechanics of the brighter-fatter effect, that loss of charges in the center columns would then lead to a slight shift of field lines in towards them, which would manifest as an excess of charges accumulating in the edge columns as the exposure continues. A mechanism like this does not require charge conservation, and would match qualitatively with the observed evolution of dips as a function of exposure time, but it is also missing an adequate description of the charge loss across potential barriers.

### 5.2 Sensor variability

The variability of the "dark dips" phenomenon across sensors that were produced using the same design and manufacturing process is quite remarkable. We looked for correlations with other known variations within the same sensors.
For instance, a number of ITL sensors were affected by an issue of excessive injection of the serial clock signals into the output transistor [6]. The most affected sensors were selected out when building the LSST focal plane. However, as seen in noise maps (Figure 3 in [1]), a few remain that exhibit the typical noise pattern associated with this issue, which is that the center segments have higher noise than the edge segments: R01_S21, R03_S01, R42_S02, and R43_S02. There is no correlation with the "dark dips" issue, as R01_S21 is among the most affected, R42_S02 slightly less, but R03_S01 and R43_S02 are not at all affected.
There is also a strong variability of the full well among ITL sensors, as measured by various estimators (see Figure 25 in [2]). There seems to be no correlation with the "dark dips" issue, as sensors at both ends of the full-well range can exhibit similar levels of dips: for instance, R01_S10 (linearity turn-off around 80,000 e-) and R01_S11 (linearity turn-off around 125,000 e-) are both strongly affected, R03_S20 (linearity turn-off around 70,000 e-) and R02_S00 (linearity turn-off around 120,000 e-) are both affected only at high saturated counts (above 1,000).
The values of the brighter-fatter coefficients $a_{00}$, $a_{01}$ and $a_{10}$ [4] could also be a probe into the strength of the sensor potential barriers, but measurements of these coefficients over all ITL sensors [2] showed no strong variations that would be correlated with dips presence.
Lastly, variability among sensors that are designed and operated in the exact same ways suggests some variation during the manufacturing process, either in the properties of the high-resistivity substrate, or during the doping and deposition phases. The ITL sensors of the LSST Camera were manufactured across a number of lots, each made up of a number of wafers, and each wafer divided into four sensor dies. After sensor selection, we have still 17 pairs of sensors that have been cut from the exact same wafers in the main LSST Camera (including one case in which three sensors are cut from the same wafer), and two more pairs in ComCam. Out of those 19 pairs, 16 are very similarly affected by the "dark dips" across the whole scale, from 5 pairs that are not at all affected to 2 that are strongly sensitive, and including the three sensors that were cut from the same wafer. But that leaves us with 3 counter-examples which, if our mapping has been tracked correctly, would indicate that the cause of the "dark dips" can be non-uniform over a single wafer.

### 5.3 Mitigation strategy

Due to the complexity of the dip growth over time and the unclear process through which the image is distorted, there is no pathway to correct the "dark dips", unlike for instance the brighter-fatter effect. The solution that has been implemented in the LSST ISR pipeline is to mask whole columns with the 'ITL_DIP' flag. To avoid time-consuming processing, and the risk of missing detections, this mask is not applied when a dip is actually detected, but when the cluster of saturated pixels is above a certain count. The thresholds that determine when the mask is applied are defined on a per-sensor basis, after sorting out the sensors in broad categories. For the LSST Camera, these parameters are stored in [8] along with other configuration parameters of the pipeline.

## 6. Conclusion

The “dark dips” phenomenon, while visually striking in the worst cases, affects only a small number of ITL sensors of the LSST Camera in this manner. Aside from these few sensors, the mitigation strategy through masking whole columns will only be necessary when extremely bright objects are exposed, which would already require masking of a significant portion of that area. Our studies allow us to tailor the masks to each sensor, and thus limit unnecessary loss of pixels.

With regards to a physical explanation of the phenomenon, we can only present the empirical observations that we have, and no definitive theory at this time. More extensive studies, using the trove of on-sky data that is becoming available, could give additional hints. For instance, we could get a much more granular view of the spatial distribution of the dip onsets and of the affected columns. Accumulating data in all filters could also give us the precision required to reveal wavelength dependency similar to the brighter-fatter effect, which would clarify at least some of the mechanisms at play in the “dark dips” phenomenon.

## Acknowledgements

This material is based upon work supported in part by the National Science Foundation through Cooperative Agreement AST-1258333 and Cooperative Support Agreement AST-1202910 managed by the Association of Universities for Research in Astronomy (AURA), and the Department of Energy under Contract No. DE-AC02-76SF00515 with the SLAC National Accelerator Laboratory managed by Stanford University. Additional Rubin Observatory funding comes from private donations, grants to universities, and in-kind support from LSST-DA Institutional Members. This work has been supported by the French National Institute of Nuclear and Particle Physics (IN2P3) through dedicated funding provided by the National Center for Scientific Research (CNRS). This work has been supported by STFC funding for UK participation in LSST, through grant ST/Y00292X/1.
This publication is based in part on proprietary Rubin Observatory Legacy Survey of Space and Time (LSST) data, and was prepared in accordance with the Rubin Observatory data rights and access policies. All authors of this publication meet the requirements for co-authorship of proprietary LSST data.
This research uses services or data provided by the Rubin Science Platform at NSF-DOE Vera C. Rubin Observatory, which is jointly funded by the U.S. National Science Foundation and the U.S. Department of Energy, Office of Science.